\documentclass[twocolumn]{aastex61}

\usepackage{amsmath}
\usepackage{color}
\usepackage{ulem}

\def\ltsima{$\; \buildrel < \over \sim \;$}
\def\simlt{\lower.5ex\hbox{\ltsima}}
\def\gtsima{$\; \buildrel > \over \sim \;$}
\def\simgt{\lower.5ex\hbox{\gtsima}}

\newcommand{\ASAGAO}{ASAGAO}
\newcommand{\ttsa}{26 arcmin$^2$}
\newcommand{\lir}{$L_{\rm IR}$}
\newcommand{\lx}{$L_{\rm X}$}
\newcommand{\nh}{$N_{\rm H}$}
\newcommand{\lognh}{$\log N_{\rm H}$}
\newcommand{\solarmass}{\mbox{${M_{\odot}}$}}

\newcommand{\solarlum}{\mbox{${L_{\odot}}$}}
\newcommand{\etal}{{\it et al.}}

\newcommand{\ergs}{erg cm$^{-2}$ s$^{-1}$}
\newcommand{\erg}{erg s$^{-1}$}

\lefthead{Ueda et al.}
\righthead{AGN Properties of Millimeter Galaxies in GOODS-S}
\begin{document}

\title{
ALMA Twenty-six arcmin$^2$ Survey of GOODS-S at One-millimeter (ASAGAO):
X-ray AGN Properties of Millimeter-Selected Galaxies
}

\author{Y.~Ueda}
\affiliation{Department of Astronomy, Kyoto University, Kyoto 606-8502, Japan}

\author{B.~Hatsukade}
\affiliation{Institute of Astronomy, The University of Tokyo, 2-21-1 Osawa, Mitaka, Tokyo 181-0015, Japan}

\author{K.~Kohno}
\affiliation{Institute of Astronomy, The University of Tokyo, 2-21-1 Osawa, Mitaka, Tokyo 181-0015, Japan}
\affiliation{Research Center for the Early Universe, The University of
Tokyo, 7-3-1 Hongo, Bunkyo-ku, Tokyo 113-0033, Japan}

\author{Y.~Yamaguchi}
\affiliation{Institute of Astronomy, The University of Tokyo, 2-21-1 Osawa, Mitaka, Tokyo 181-0015, Japan}

\author{Y.~Tamura}
\affiliation{Department of Physics, Nagoya University, Furo-cho, Chikusa-ku, Nagoya 464-8601, Japan}

\author{H.~Umehata}
\affiliation{The Open University of Japan, 2-11, Wakaba, Mihama-ku, Chiba, Chiba 261-8586, Japan}

\author{M.~Akiyama}
\affiliation{Astronomical Institute, Tohoku University, 6-3 Aramaki, Aoba-ku, Sendai 980-8578, Japan}

\author{Y.~Ao}
\affiliation{National Astronomical Observatory of Japan, 2-21-1 Osawa, Mitaka, Tokyo 181-8588, Japan}

\author{I.~Aretxaga}
\affiliation{Instituto Nacional de Astrof\'isica, \'Optica y Electr\'onica (INAOE), Aptdo. Postal 51 y 216, 72000 Puebla, Pue., Mexico}

\author{K.~Caputi}
\affiliation{
Kapteyn Astronomical Institute, University of Groningen, 9700 AV
Groningen, The Netherlands}

\author{J.\,S.~Dunlop}
\affiliation{Institute for Astronomy, University of Edinburgh, Royal Observatory, Blackford Hill, Edinburgh EH9 3HJ, UK}

\author{D.~Espada}
\affiliation{National Astronomical Observatory of Japan, 2-21-1 Osawa, Mitaka, Tokyo 181-8588, Japan}
\affiliation{The Graduate University for Advanced Studies
(SOKENDAI), 2-21-1 Osawa, Mitaka, Tokyo, 181-8588, Japan}

\author{S.~Fujimoto}
\affiliation{Institute for Cosmic Ray Research, The University of Tokyo, Kashiwa, Chiba 277-8582, Japan}

\author{N.\,H.~Hayatsu}
\affiliation{Department of Physics, The University of Tokyo, 7-3-1 Hongo, Bunkyo-ku, Tokyo 113-0033, Japan}

\author{M.~Imanishi}
\affiliation{National Astronomical Observatory of Japan, 2-21-1 Osawa, Mitaka, Tokyo 181-8588, Japan}

\author{A.\,K.~Inoue}
\affiliation{Department of Environmental Science and Technology,
Faculty of Design Technology, Osaka Sangyo University, 3-1-1, Nakagaito, Daito, Osaka 574-8530, Japan}

\author{R.\,J.~Ivison}
\affiliation{European Southern Observatory, Karl-Schwarzschild-Str.~2,
D-85748 Garching, Germany}
\affiliation{Institute for Astronomy, University of Edinburgh, Royal Observatory, Blackford Hill, Edinburgh EH9 3HJ}

\author{T.~Kodama}
\affiliation{Astronomical Institute, Tohoku University, 6-3 Aramaki, Aoba-ku, Sendai 980-8578, Japan}

\author{M.\,M.~Lee}
\affiliation{Department of Astronomy, The University of Tokyo, 7-3-1 Hongo, Bunkyo-ku, Tokyo 113-0033, Japan}
\affiliation{National Astronomical Observatory of Japan, 2-21-1 Osawa, Mitaka, Tokyo 181-8588, Japan}

\author{K.~Matsuoka}
\affiliation{
Dipartimento di Fisica e Astronomia, Universit\`a di Firenze
Via G. Sansone 1, I-50019 Sesto Fiorentino (Firenze), Italy
}

\author{T.~Miyaji}
\affiliation{Instituto de Astronom\'{i}a, Universidad Nacional
Aut\'{o}noma de M\'{e}xico, Ensenada, Baja California, Mexico; P.O.\ Box
439027, San Diego, CA 92143-9027, USA}

\author{K.~Morokuma-Matsui}
\affiliation{Institute of Space and Astronautical Science, Japan Aerospace Exploration Agency, 3-1-1 Yoshinodai, Chuo-ku, Sagamihara, Kanagawa 252-5210, Japan}

\author{T.~Nagao}
\affiliation{Research Center for Space and Cosmic Evolution, Ehime University,
2-5 Bunkyo-cho, Matsuyama, Ehime 790-8577, Japan}

\author{K.~Nakanishi}
\affiliation{National Astronomical Observatory of Japan, 2-21-1 Osawa, Mitaka, Tokyo 181-8588, Japan}
\affiliation{The Graduate University for Advanced Studies
(SOKENDAI), 2-21-1 Osawa, Mitaka, Tokyo, 181-8588, Japan}

\author{K.~Nyland}
\affiliation{National Radio Astronomy Observatory, 520 Edgemont Rd., Charlottesville, VA 22903, USA}

\author{K.~Ohta}
\affiliation{Department of Astronomy, Kyoto University, Kyoto 606-8502, Japan}

\author{M.~Ouchi}
\affiliation{Institute for Cosmic Ray Research, The University of Tokyo, Kashiwa, Chiba 277-8582, Japan}

\author{W.~Rujopakarn}
\affiliation{Department of Physics, Faculty of Science,
Chulalongkorn University, 254 Phayathai Road, Pathumwan, Bangkok 10330, Thailand}
\affiliation{National Astronomical Research Institute of Thailand (Public Organization), Donkaew, Maerim, Chiangmai 50180, Thailand}
\affiliation{Kavli Institute for the Physics and Mathematics of the universe (WPI), The University of Tokyo Institutes for Advanced Study, The University of Tokyo, Kashiwa, Chiba 277-8583, Japan}

\author{T.~Saito}
\affiliation{Nishi-Harima Astronomical Observatory, Centre of Astronomy,
University of Hyogo, 407-2 Nishigaichi, Sayo, Sayo-gun, 679-5313 Hyogo, Japan}

\author{K.~Tadaki}
\affiliation{National Astronomical Observatory of Japan, 2-21-1 Osawa, Mitaka, Tokyo 181-8588, Japan}

\author{I.~Tanaka}
\affiliation{Subaru Telescope, National Astronomical Observatory of Japan}

\author{Y.~Taniguchi}
\affiliation{The Open University of Japan, 2-11, Wakaba, Mihama-ku, Chiba, Chiba 261-8586, Japan}

\author{T.~Wang}
\affiliation{Institute of Astronomy, The University of Tokyo, 2-21-1 Osawa, Mitaka, Tokyo 181-0015, Japan}
\affiliation{National Astronomical Observatory of Japan, 2-21-1 Osawa, Mitaka, Tokyo 181-8588, Japan}

\author{W.-H.~Wang}
\affiliation{Institute of Astronomy and Astrophysics, Academia Sinica, Taipei 10617, Taiwan}

\author{Y.~Yoshimura}
\affiliation{Department of Astronomy, The University of Tokyo, 7-3-1 Hongo, Bunkyo-ku, Tokyo 113-0033, Japan}

\author{M.\,S.~Yun}
\affiliation{Department of Astronomy, University of Massachusetts, Amherst, MA 01003, USA}

\begin{abstract}

We investigate the X-ray active galactic nucleus (AGN) properties of
millimeter galaxies in the Great Observatories Origins Deep Survey
South (GOODS-S) field detected with the Atacama Large
Millimeter/submillimeter Array (ALMA), by utilizing the Chandra 7-Ms
data, the deepest X-ray survey to date. Our millimeter galaxy sample
comes from the ASAGAO survey covering 26 arcmin$^2$ (12 sources at a
1.2-mm flux-density limit of $\approx$0.6 mJy), supplemented by the
deeper but narrower 1.3-mm survey of a part of the \ASAGAO\ field by
Dunlop et al.\ (2017). Fourteen out of the total 25 millimeter
galaxies have Chandra counterparts. The observed AGN fractions at
$z=1.5-3$ is found to be 90$^{+8}_{-19}$\% and $57^{+23}_{-25}$\% for
the ultra/luminous infrared galaxies with $\log$ \lir/\solarlum =
12--12.8 and $\log$ \lir/\solarlum = 11.5--12, respectively. The
majority ($\sim$2/3) of the ALMA and/or Herschel detected X-ray AGNs
at $z=1.5-3$ appear to be
star-formation dominant populations, having \lx / \lir\ ratios smaller
than the ``simultaneous evolution'' value expected from the local
black-hole mass to stellar mass ($M_{\rm BH}$-$M_*$) relation. On the
basis of the \lx\ and stellar mass relation, we infer that a large
fraction of star-forming galaxies at $z=1.5-3$ have black hole masses
smaller than those expected from the local $M_{\rm BH}$-$M_*$
relation. This is opposite to previous reports on luminous AGN at same
redshifts detected in wider and shallower surveys, which are subject
to selection biases against lower luminosity AGN. Our results
are consistent with an evolutionary scenario that star
formation occurs first, and an AGN-dominant phase follows later, in
objects finally evolving into galaxies with classical bulges.

\end{abstract}


\section{Introduction}

A key issue in cosmic evolution is the growth history of supermassive
black holes (SMBHs) in galactic centers and their stellar populations,
leading to the tight bulge-mass to SMBH-mass correlation observed in
the present universe (see \citealt{kor13} for a recent review; following
them, we use the term ``bulge'' only for classical bulges and elliptical
galaxies). Overall good agreement between the star formation and mass
accretion history from $z\sim3$ to $z\sim0$ \citep[e.g.,][]{boy98,
mar04, ued15, air15} implies that they seem to have mostly
``co-evolved'' on cosmological timescales. It is not yet clear, however,
how strictly stars and the SMBH were coeval in an individual
galaxy. Large dispersion in the bulge to SMBH mass ratio in high
redshift objects suggests that the evolution processes are more complex
than a simple ``co-evolution'' hypothesis and that the observational
result may be largely subject to selection biases of the sample studied
\citep{kor13}.

Both star formation and mass accretion co-moving densities peak at
$z\sim2$ \cite[][]{mad14,ued14}, which is often referred as
``cosmic noon''. 
Submillimeter observations discovered IR luminous galaxies 
(submillimeter galaxies; SMGs) at these redshifts, where violent star
formation deeply enshrouded by dust takes place
\citep[e.g.,][]{sma97}. These objects would be a key population for
understanding the origin of galaxy-SMBH co-evolution; many theoretical
studies suggest that major mergers trigger both star formation and mass
accretion, making them appear as IR luminous galaxies containing an
obscured AGN \citep[e.g.,][]{hop06}. Sensitive X-ray observations
provide an efficient way for detecting such AGN (e.g.,
\citealt{ale05}).  Due to the limited angular resolution of previous
(single-dish) sub/millimeter or FIR observatories ($10-20''$), however,
it is often difficult to robustly determine their multiwavelength
counterparts and thereby estimate their star formation rate and mass
accretion rate.

The Atacama Large Millimeter/submillimeter Array (ALMA) is changing the
situation thanks to its unprecedented angular resolution and sensitivity
at sub/millimeter wavelengths. The GOODS-S region is one of the best
fields for studying faint AGN in distant galaxies, because the
deepest X-ray survey to date has been performed with the 
Chandra observatory there
\citep{luo17}. On the basis of an ALMA follow-up of SMGs in the 
Extended-Chandra Deep Field South (E-CDFS)
field (ALMA LABOCA E-CDFS Submillimeter
Survey, ALESS; \citealt{hod13}), \citet{wan13} determined the AGN
fraction to be 15$^{+15}_{-6}$\% at an X-ray flux limit of
$2.7\times10^{-16}$ \ergs\ (0.5--8 keV); hereafter we refer to
these AGN as the ``ALESS-AGN sample''. 
More recently, utilizing the Chandra 4 Ms and Karl G.\ Jansky Very Large
Array (JVLA) survey catalogs,
\cite{ruj16} have
identified 6 AGN from 16 millimeter galaxies
in the GOODS-S/ultra deep
field (UDF) detected by \citet{dun17} (hereafter D17), 
two of which were first discovered 
in the ALMA spectroscopic survey covering a 1 arcmin$^2$ area in the UDF 
(ASPECS, \citealt{ara16}).
Using ALMA in cycle 3, our
team has performed an unbiased deep 1.2-mm imaging survey
over a 26 arcmin$^2$ region
inside the GOODS-S-JVLA field (Alma twenty-Six Arcmin$^2$ survey of
Goods-south At One-millimeter, ASAGAO), 
which fully encompasses the
GOODS-S/UDF.  This survey fills the gap in the survey parameter space
(sensitivity and area) between the ALESS and the UDF survey.  The
results using the JVLA radio-continuum data, which are also important
for identifying AGN, will be reported in a separate paper (Rujopakarn
et al.\ in prep.).

In this paper, we investigate the X-ray AGN properties of millimeter
galaxies detected in our survey and that by D17, utilizing the latest
Chandra 7 Ms source catalog \citep{luo17}. We mainly refer to the
FourStar Galaxy Evolution Survey (ZFOURGE) catalog for identification of
ALMA sources detected in the \ASAGAO\ (the source catalog will be presented
in Hatsukade \etal, in prep.). The ZFOURGE catalog also provides a
sample of galaxies not detected with ALMA located in our survey
region. We investigate relations among X-ray luminosity (or mass
accretion rate), infrared luminosity (or star formation rate, SFR), and
stellar mass of these samples, and discuss the implications on the
galaxy-SMBH co-evolution at cosmic noon. 
Throughout the paper, we adopt the conversion from IR luminosity
(without AGN contribution) to SFR as SFR/(\solarmass\ yr$^{-1}$) $=
1.09\times10^{-10} L_{\rm IR}/\solarlum$, which is based on
\citet{ken98} re-calibrated for a \citet{cha03} initial mass function
(IMF)\footnote{Any SFR values quoted from the literature are re-scaled
to this calibration.}.  The cosmological parameters of $H_{0} = 70$ km
s$^{-1}$ Mpc$^{-1}$, $\Omega_{\rm M}=0.3$, and $\Omega_{\Lambda}=0.7$
are adopted. The errors in the number fraction based on a small-size
sample ($\leq10$) are given at 1$\sigma$ confidence limits referring to
Table~6 of \citet{geh86}.

\section{Millimeter Galaxy Catalog}

\subsection{Observations and Source Detection}

ALMA band 6 (1.2 mm) observations of the $5' \times 5'$ ($\approx 26$
arcmin$^2$) area of the \ASAGAO\ were conducted in
September 2016 in the C40-6 array configuration for a total observing
time of 45 hours (Project code: 2015.1.00098.S, PI: K. Kohno). The
details of the observations and data reduction are described in
Hatsukade \etal\, (in prep.), and here we briefly summarize them. Two
frequency tunings were adopted centered at 1.14~mm and 1.18 mm to cover
a wider frequency range, whose central wavelength was 1.16 mm. The
correlator was used in the time domain mode with a bandwidth of 2000 MHz
(15.625 MHz $\times$ 128 channels). Four basebands were used for each
tuning, and the total bandwidth was 16 GHz
covering the 244--248 GHz, 253--257 GHz, 259--263 GHz, and
268--272 GHz frequency ranges.
The number of available antennas was 38--45.

The data were reduced with Common Astronomy Software Applications
\citep[CASA;][]{mcmu07}. The maps were processed with
the {\bf CLEAN} algorithm with the task 
{\bf tclean}\footnote{The adopted parameters are as follows: natural weighting, cell size of 0.15
arcsec, gridder of mosaic, specmode of muti-frequency synthesis, and
nterms of 2.}.
Clean boxes were placed when a component with a peak signal-to-noise
ratio (SN) above 5.0 is identified, and
{\bf CLEAN}ed down to a $2\sigma$ level. The observations were done
with a higher angular resolution ($\sim$$0\farcs2$) than originally
requested ($0\farcs8$), and we adopted a $uv$-taper of 160~k$\lambda$ to
improve the surface brightness sensitivity, which gives the final
synthesized beamsize of $0\farcs94 \times 0\farcs67$ and the typical rms
noise level of 89~$\mu$Jy~beam$^{-1}$.

Source detection was conducted on the image before correcting for the
primary beam attenuation. The source and noise properties were estimated with
the {\sc Aegean} \citep{hanc12} source-finding algorithm. We find 12
sources with a peak SN threshold of $5.0\sigma$, whereas no negative
source with a peak SN $< -5.0\sigma$ is found. The integrated flux
density is calculated with elliptical Gaussian fitting. Table~1 (2--4th
columns) lists the position of the peak intensity and the integrated
flux density corrected for the primary beam attenuation with its
$1\sigma$ error.

\subsection{Definition of Our ALMA Sample}

To supplement our \ASAGAO\ sample, we also include fainter flux ALMA sources
detected in the deep 1.3 mm image of the UDF ($\simeq$4.5 arcmin$^2$
located inside the \ASAGAO\ field) by D17.  By
position matching ($<0\farcs2$) and flux comparison, we find that the
objects with IDs 3, 6, and 8 are identical to UDF1, UDF2, UDF3 in D17,
respectively. 
Table~1 (5th column) lists the 1.3 mm flux density of the D17
sources. 
We refer to D17 for the positions of these sources
except UDF1, UDF2, and UDF3 (2nd--3rd columns of Table~1).
In addition, we find that ID 8 (UDF3) and UDF8 are also detected in the
1.2-mm continuum map of the ASPECS \citep{ara16}. The 1.2 mm fluxes obtained by
ASPECS are listed in the 6th column of Table~1 for these sources. 
The fluxes of UDF3 obtained from \ASAGAO\ and ASPECS are consistent 
within the errors.

The \ASAGAO\ sources are cross-matched against the
ZFOURGE catalog \citep{str16}, after correcting for systematic
astrometric offsets ($-0\farcs09$ in RA and $+0\farcs28$ in Dec) with respect
to the ALMA image, which are calibrated by the positions of stars 
in the Gaia Data Release 1 catalog \citep{gai16} within the \ASAGAO\ field.
The ZFOURGE fully covers the \ASAGAO\ \ttsa\ field, in
which $\sim$3,000 objects are cataloged with limiting magnitudes of
$K_{\rm S}$ (AB) = 26.0 to 26.3 (5$\sigma$) at the 80\% and 50\%
completeness levels with masking, respectively. We search for counterparts from the ZFOURGE
catalog whose angular separation from the ALMA position is smaller than
$0\farcs2$, corresponding to $\approx 3\sigma$ of the statistical
positional error of ALMA for a point-like source; when a counterpart is
largely extended (IDs 3 and 5), we allow a larger positional offset, up to 1
arcsec. Consequently, ZFOURGE counterparts are found for 10 sources,
except for IDs 9 and 11. Since the chance probability that an unrelated
object falls within a radius of $0\farcs2$ is only $\approx$0.004 as
calculated from the source density of the ZFOURGE catalog, we can
safely assume that all of these ALMA-ZFOURGE associations are true
\footnote{Some ZFOURGE objects may be lensing galaxies.}.
We adopt the best-estimated redshift (spectroscopic or photometric) in the
ZFOURGE catalog but refer to D17 for the UDF sample and IDs 3, 6, and 8. 
The adopted redshift is listed in the 14th column of Table~1.
The differences between the photometric redshifts in ZFOURGE catalog and
the spectroscopic redshifts in D17 are found to be $\Delta z/(1+z) <
0.08$ with a median of 0.02. This indicates that the photometric
redshift errors little affect our analysis and conclusions.

We regard the 23 sources with ZFOURGE counterparts (10 sources from the
\ASAGAO\ excluding IDs 9 and 11, and 13 sources from D17 excluding the
overlapping sources UDF1, UDF2, and UDF3; hereafter the \ASAGAO\ sample
and the UDF sample, respectively) as the parent sample for our
subsequent studies. All of them are also detected in FIR bands (70--160
$\mu$m) with Herschel/PACS \citep{elb11}. 

\subsection{SED fitting and Infrared Luminosities}

To estimate the infrared luminosities (and hence SFRs) of the ASAGAO sample,
we analyze their spectral energy distribution
(SED), utilizing the MAGPHYS code \citep{dacunha2008,dacunha2015}.
We use 43 optical-to-millimeter photometric data including the ALMA,
ZFOURGE, and de-blended Herschel/SPIRE photometries (Wang, T., et al. in
prep.). Checking the ALMA spectra, we have confirmed that line
contamination to the 1.2 mm continuum flux is ignorable in all the
targets. We use MAGPHYS high-z excitation code for the sources at $z>1$,
whereas the MAGPHYS original package is applied for ID 5 ($z=0.523$). In
the SED fitting, the redshift is fixed to the value in Table~1. 
The resultant infrared luminosities in the rest-frame 8--1000 
$\mu$m band (\lir )\footnote{The MAGPHYS code returns the 3--1000
$\mu$m luminosities but we adopt these values as the conventional
8--1000 $\mu$m band luminosities for the ASAGAO
sample because the contribution of the 3--8 $\mu$m band to 
total dust emission is negligible \citep{cle13}.} 
are listed in the 7th column of Table 1.
Further details of the SED analysis are given in
Yamaguchi et al.\ (in prep.).

The infrared luminosity of the ASAGAO sample we derive 
with the MAGPHYS code
are found to be consistent with
the \lir\ values 
in D17 (for IDs 3, 6, and 8) and 
in \citet{str16} within 0.1--0.2 dex
in most cases; the maximum difference of $\approx$0.5 dex is found
for ID~8 (UDF3) between our result and D17, whose 1.3 mm flux density
was corrected for unusually large line emission in the analysis of D17.
\citet{str16} obtained the infrared luminosities
by fitting the 24, 100, and 160 $\mu$m
photometry with the \citet{wuy08} template.
For the UDF sample
we refer to D17 for \lir \footnote{Following the recipe in Section~6.3 of
D17, we converted the SFR$_{\rm FIR1}$ values in their Table~4 into
\lir\ using the \citet{mur11} relation with a minor correction from a
Chabrier to a Kruopa IMF.}, which were 
obtained by a SED fit to 24 $\mu$m to
1.3 mm photometry with the spectral template by \citet{kir15}.
The listed \lir\ values include an estimated 20\% AGN contribution.

The flux densities of the \ASAGAO\ sample range from $\sim$0.6 to
$\sim$3 mJy. It bridges the ALESS sample \citep{hod13}
\footnote{By transforming the 850 $\mu$m flux densities to
the 1.2 mm ones, the ALESS fluxes reach a similar depth ($\sim 1$mJy 
) to that of the ASAGAO survey. The major difference is that 
the ALESS is pointed to bright SMGs, while the ASAGAO is a blind survey.}
and the UDF sample (D17), which contain brighter
and fainter sub/millimeter galaxies than the \ASAGAO\ sources,
respectively.
The majority of our sources with ZFOURGE counterparts are located at $z
\simeq 2-3$.  No object at $z>3$ has been identified, although it
is possible that the two sources (IDs 9 and 11) lacking redshift
constraints are $z>3$ galaxies. This result is consistent with the
findings by \citet{ara16} and D17 that high-z galaxies are rare in the
faint ALMA populations, as expected on the basis of the selection
wavelength and depth of the survey \citep{bet15}. Eight out of the
10 identified \ASAGAO\ sources have infrared luminosities larger than
$10^{12}$ \solarlum\ and hence are classified as 
``ultra-luminous infrared galaxies'' (ULIRGs).
The star forming rate (SFR) estimated
from the infrared luminosity after subtracting an estimated 20\% AGN
contribution (D17) ranges 9--600 \solarmass\ yr$^{-1}$.

\section{Cross Matching with Chandra 7 Ms Catalog}

We cross-match the ALMA source list with the Chandra 7 Ms source catalog
of the CDFS \citep{luo17}, which contains 1008 objects 
detected in
0.5--2 keV, 2--7 keV, and/or 0.5--7 keV bands,
among which 137 objects are located within the ASAGAO field.
The sensitivity limit of
the Chandra data is $(2-8)\times10^{-17}$ \ergs\ (0.5--7 keV) for a power
law photon index of 1.4, which is not uniform in our ALMA survey
region. A flux of $2\times10^{-17}$ \ergs\ (0.5--7 keV)
corresponds to an intrinsic 0.5--8 keV
luminosity of $\log L_{\rm X}$ / (erg s$^{-1}$) = 41.7 
and 41.9 for a source at $z=2$ 
with an absorption of $\log N_{\rm H} / {\rm cm}^{-2}$ = 20 and 23,
respectively, assuming our model spectrum 
with a photon index of 1.9 (see below).
X-ray counterparts
are selected if the angular separation between the ALMA and Chandra
sources is smaller than 3 times their combined $1\sigma$ positional
error, which is dominated by the Chandra one (0.14--0.62 arcsec,
depending on the photon counts). We found 14 X-ray
counterparts to the 23 ALMA sources, 8 from the \ASAGAO\ sample (10
objects) and 6 from the UDF sample (13 objects). We confirm that all 
6 of the X-ray sources in the D17 sample reported by \citet{ruj16} using the
shallower Chandra 4 Ms catalog \citep{xue11} are included. The
probability of spurious identification is negligibly small, $<$0.01,
which is estimated from the surface number density of the Chandra
sources in our ALMA survey region ($\sim 2\times10^4$ deg$^{-2}$).

Following the recipe described in \citet{ued03}, we estimate the column
density and intrinsic (de-absorbed) luminosity of each Chandra
object. As the model spectrum, we assume a cutoff power law spectrum
plus its reflection component from optically-thick cold matter with a
solid-angle of 2$\pi$. Both the cutoff power law and its reflection
are subject to intrinsic absorption at the
source redshift and Galactic absorption, which is fixed at \nh\ =
$8.8\times10^{19}$ cm$^{-2}$. Such a reflection component from the torus
and/or the accretion disk 
is known to be commonly present in the X-ray spectra of AGN (e.g.,
\citealt{kaw16}). From the hardness ratio of the vignetting corrected count
rates in the 0.5--2 keV and 2--7 keV band, we first determine the
apparent photon index by assuming no intrinsic absorption. If it is
found to be larger than 1.9, then we adopt this value and consider no
absorption. Otherwise, we fix the intrinsic photon index to 1.9 and
determine the absorption to account for the observed hardness ratio. For
objects detected only in the total (0.5--7 keV) band, we assume a photon
index of 1.9 and an absorption column density of \lognh\ / cm$^{-2}$ =
20. Then, the intrinsic luminosity is calculated based on the photon
index and normalization. The statistical error in the hardness ratio is
taken into account to estimate the uncertainties in the photon index,
absorption, and intrinsic luminosity. The results are listed in the
11--13th columns of Table~1. 

The X-ray detection rate of our ALMA sample is $61\pm10$\% (14 out of
23), and $80^{+13}_{-20}$\% for the \ASAGAO\ sample (8 out of 10). We
adopt similar criteria proposed by \citet{wan13} to classify an X-ray
detected object as an AGN: (I) the effective photon index ($\Gamma_{\rm
eff}$ ) is smaller than 1.0 suggestive of intrinsic absorption, (II) the
de-absorbed rest-frame 0.5--8 keV luminosity ($L_{\rm X}$) is larger
than $3\times10^{42}$ \erg , (III) the apparent rest-frame 0.5--8 keV
luminosity ($L_{\rm X}^{\rm app}$) is higher than 5 times that of star
formation ($L_{\rm X, SF}$) estimated from the infrared luminosity
according to the formula of \citet{leh10} \footnote{$\log (L_{\rm
X,SF}/1.21) = 39.49 + 0.74\times \log( 9.8\times10^{-11}L_{\rm
IR}/\solarlum)$}, and (IV) the observed X-ray to MIR flux ratio is $\log
(f_{\rm X}/f_{\rm 3.6 \mu m}) > -1$.  We then regard sources that
satisfy at least one criterion (the 15th column of Table~1) and have $\log
L_{\rm X} > 41.5$ are AGN. The 13 ALMA-Chandra objects
except UDF9 are classified as AGN.

For the 9 ALMA sources that are not detected with Chandra, we give in
Table~1 (10th and 13th columns) a 90\% confidence upper limit of the
observed flux (converted from the count-rate upper limit in that
position by assuming a photon index of 1.4, that of the X-ray
background, with no absorption), and that of the intrinsic luminosity
obtained by assuming a photon index of 1.9 and an absorption of \lognh\
/ cm$^{-2}$ = 23 as a typical spectrum of an obscured AGN when the
redshift is known. We also perform stacking analysis of the 7 Ms Chandra
images for these 9 sources with the CSTACK 
program\footnote{http://lambic.astrosen.unam.mx/cstack/}, 
which utilizes
event files reprocessed by \citet{cap16} and \citet{wil17}.
No significant
signals are detected in the 0.5--2 keV and 2--8 keV bands with 90\%
upper limits of $5\times10^{-18}$ \ergs\ and $2\times10^{-17}$ \ergs\ 
for the mean fluxes, respectively (assuming a photon index of 1.4).

To investigate the nature of the X-ray sources in the ASAGAO field that
are not detected with ALMA, we cross-match them with the ZFOURGE catalog
in the same way as above. We find that out of the 
123 ALMA-undetected
Chandra sources, 
111 have ZFOURGE counterparts, among which 
95 are
detected Herschel/PACS and hence have estimates of their infrared
luminosities. When we limit the redshift range to $z=1.5-3$, where most
of the ALMA-Chandra sources reside, 
there are 49 Chandra sources within 
the ASAGAO field, out of which 46 have ZFOURGE counterparts.
Among them, 13 are detected with ALMA (and also with
Herschel), 
26 with Herschel only, and 
7 are neither detected with ALMA
nor with Herschel. For comparison with our ASAGAO and UDF sample, we
refer to the ALMA-undetected but Herschel-detected Chandra AGN at
$z=1.5-3$ as the ``Herschel-AGN sample'' in the subsequent discussions.

\section{Results and Discussion}

\subsection{Correlation between Star Formation Rate and Stellar Mass}

Figures~1 and 2 plot the relation between infrared luminosity and 
redshift and that between stellar mass ($M_*$) and SFR, respectively, for
the X-ray sources in the \ASAGAO\ and UDF samples (red circles and
triangles), the ALESS-AGN sample (magenta squares), and the Herschel-AGN
sample (black diamonds).
We refer to \citet{wan13} and \citet{str16} for 
the infrared luminosities of the ALESS-AGN and Herschel-AGN samples, 
respectively.
For all the samples, we estimate the SFRs from \lir\ after subtracting
an assumed 20\% AGN contribution. The stellar masses (listed in the
8th column in Table~1) are taken from the ZFOURGE catalog
\citep{str16} for the \ASAGAO\ (except IDs 3, 6, and 8) and
Herschel-AGN samples, from D17 for the UDF sample and IDs 3, 6, and 8,
and from \citet{wan13} for the
ALESS-AGN sample (re-calibrated for a Chabrier IMF).
All the $M_*$ values were derived with multiwavelength SED
analyses by assuming the solar abundances, 
and represent the total stellar masses in the galaxies.
In the figure, we draw the ``main sequence'' lines at $z=2.75$,
$z=2.25$, and $z=1.75$ according to \citet{spe14}.  Figure~1 indicates
that most of the ALMA-Chandra AGN belong to main-sequence star-forming
galaxies (SFGs), although 
a few ASAGAO objects are classified as starburst
galaxies, being located more than 0.6 dex above the main-sequence
line (\citealt{rod11}).
This is consistent
with previous ALMA survey results for faint sub/millimeter galaxies
\citep[e.g.,][]{hat15,yam16}. The ALMA undetected AGN generally follow
the same correlation at lower stellar-mass or SFR ranges.

\begin{figure}
\epsscale{0.45}
\includegraphics[angle=0,scale=0.8]{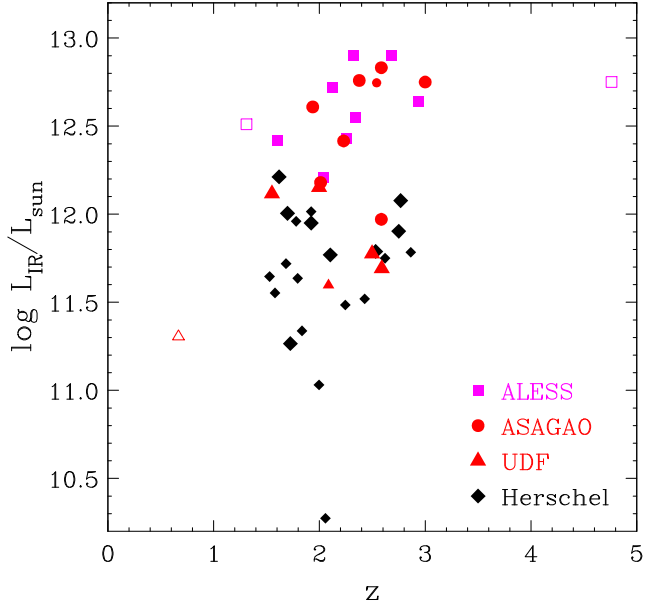}
\caption{
Relation between redshift ($z$) and infrared luminosity in the 
rest-frame 8--1000 $\mu$m band (\lir) for various samples. 
Red filled circles: 
Chandra detected sources in the \ASAGAO\ sample
that are classified as AGN at $z=1.5-3$.
Red filled triangles:
Chandra detected sources in the UDF sample (D17)
that are classified as AGN at $z=1.5-3$.
Red open triangle: UDF9 (not AGN).
Magenta filled squares:
the ALESS-AGN sample \citep{wan13} at $z=1.5-3$.
Magenta open squares: those not at $z=1.5-3$.
Black diamonds: 
the Herschel-AGN sample ($z=1.5-3$).
Smaller symbols correspond to those with stellar masses of $\log M_* /
 \solarmass < 10.5$.
\label{fig1}}
\end{figure}

\begin{figure}
\epsscale{0.45}
\includegraphics[angle=0,scale=0.8]{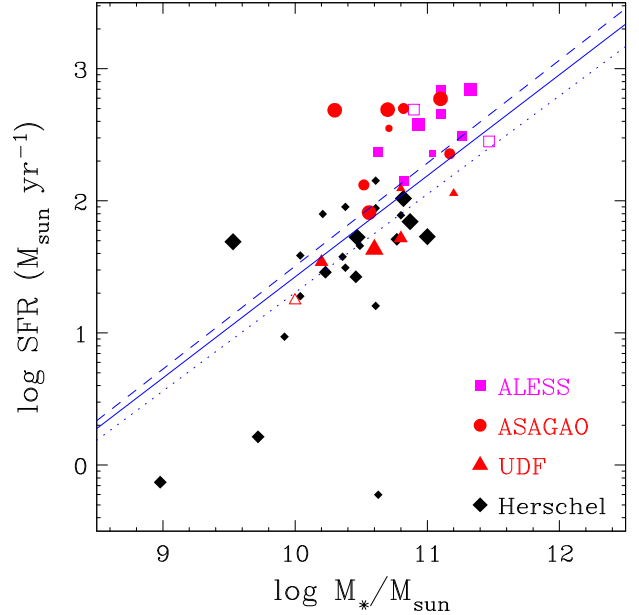}
\caption{
Correlation between stellar mass and star forming rate (SFR).
Red filled circles: 
Chandra detected sources in the \ASAGAO\ sample
that are classified as AGN at $z=1.5-3$.
Red filled triangles:
Chandra detected sources in the UDF sample (D17)
that are classified as AGN at $z=1.5-3$.
Red open triangle: UDF9 (not AGN).
Magenta filled squares:
the ALESS-AGN sample \citep{wan13} at $z=1.5-3$.
Magenta open squares: those not at $z=1.5-3$.
Black diamonds: 
the Herschel-AGN sample ($z=1.5-3$).
Large, medium, and small symbols correspond to objects at $z=2.5-3$,
 $z=2-2.5$ and $z=1.5-2$, respectively. 
The blue lines denote the main sequence relations at $z=2.75$ (dashed) ,
 $z=2.25$ (solid), and $z=1.75$ (dotted), according to \citet{spe14}.
\label{fig2}}
\end{figure}

\subsection{AGN Fraction}

We derive the AGN fraction in the millimeter galaxies down to an
X-ray flux limit of $\sim 5\times10^{-17}$ \ergs\ (0.5--7 keV), which is
5 times fainter than that of the ALESS-AGN sample \citep{wan13}. Following
\citet{wan13}, we calculate the cumulative AGN fraction at $f_{\rm X} >
f_{\rm X, lim}$ as $ \sum_{i=1}^{N} (1/N_{{\rm MG},i})$, where the
suffix $i$ (1 through $N$) represents each Chandra-identified AGN with a
flux of $f_{\rm X, i} (> f_{\rm X, lim})$ and $N_{{\rm MG},i}$ is the
total number of ALMA objects that would be detected if they had fluxes
brighter than $f_{\rm X, i}$. We find that the AGN fraction in the \ASAGAO\
sample (with a flux-density limit of $>$0.6 mJy at 1.2 mm) is
$67^{+15}_{-19}$\% at $f_{\rm X} > 5\times10^{-17}$ \ergs\ (0.5--7 keV),
whereas that in the UDF sample covering a flux density range of
0.15--0.31 mJy at 1.3 mm (or 0.18--0.39 mJy at 1.2 mm)
is $38^{+18}_{-15}$\% at $f_{\rm X} >
4\times10^{-17}$ \ergs\ (0.5--8 keV).
When we limit the sample to those at $z=1.5-3$, we obtain AGN fractions
of $88^{+10}_{-24}$\% in the \ASAGAO\ sample and $63^{+19}_{-24}$\% in the UDF
sample (at the same X-ray flux limits as above).

The best-estimated AGN fraction obtained from the \ASAGAO\ sample is higher
than that from the UDF sample, although the significance of the
difference is marginal due to the limited sample size. This is most
likely related to the lower SFRs (hence lower IR luminosities) or
smaller stellar masses in the latter sample. In fact, when we divide the
combined \ASAGAO +UDF sample at $z=1.5-3$ by IR luminosity, we find AGN
fractions of $90^{+8}_{-19}$\% at $f_{\rm X} > 5\times10^{-17}$ \ergs\
(0.5--7 keV) for the ULIRGs ($\log$ \lir\ = 12--12.8), and $57^{+23}_{-25}$\%
at $f_{\rm X} > 4\times10^{-17}$ \ergs\ (0.5--8 keV) for the LIRGs ($\log$
\lir\ = 11.5--12). The trend is consistent with previous results that
the AGN fraction is small among faint millimeter galaxies \citep{fuj16}
or among $z\sim2$ galaxies with small stellar masses
\citep{kri07,yam09,wan17}.

Our \ASAGAO\ result ($\sim$90\%) is even higher than that by \citet{wan17},
who obtained an AGN fraction of $\sim$50\% among their green-color
galaxy sample with $\log M_* / \solarmass > 10.6$. This may be related
to the fact that the high-resolution observations of ALMA are biased
towards more compact objects; if that is the case, these results imply
that compact SFGs have remarkably high AGN fractions, as suggested by \citet{cha17} for $z<1.5$ AGN.
On the other hand,
the ALESS sources, whose stellar masses are larger than the \ASAGAO\ sample
(Figure~1), apparently shows a much smaller AGN fraction ($\sim$20\%)
than ours. A primary reason is its brighter X-ray flux limit, because
most of the ALESS region (E-CDFS) is covered only by $\sim$250 ks
exposure of Chandra. Indeed, when we impose the same X-ray flux limit as
for the ALESS-AGN sample ($f_{\rm X} > 2.7\times10^{-16}$ \ergs ), the AGN
fraction for the \ASAGAO\ $z=1.5-3$ sample becomes $\sim$25\%, consistent
with the ALESS result. 
Namely, many AGN in these millimeter galaxies are not X-ray luminous,
and hence are only detectable with very deep X-ray data.
This trend implies that even $\sim$90\% may be a lower limit, getting higher 
when the Chandra exposure increases beyond 7 Ms. 

\subsection{X-ray Absorption Properties}

Among the 12 ALMA-Chandra sources for which the X-ray hardness ratio is
available, 7 have best-fit absorption column densities of
$\log N_{\rm H} / {\rm cm}^{-2} < 22$ (classified as ``X-ray type-1
AGN'' according to \citealt{ued03}) and 5 show $\log N_{\rm H} / {\rm
cm}^{-2} \geq 22$ (``X-ray type-2 AGN''). While the obscuration fraction (5
out of 12) is consistent with that found from hard X-ray (3--24
keV) detected U/LIRGs in the COSMOS field at $z=0.3-1.9$ (12 out of 23,
\citealt{mat17}),
it looks much smaller than those of local U/LIRGs; according to a
recent study by \citet{ric17}, more than 90\% of AGN in late-merger
galaxies are subject to heavy obscuration $\log N_{\rm H} / {\rm
cm}^{-2} > 23$. 
Since the X-ray luminosity range of our ALMA-Chandra sample is similar
to that of typical local U/LIRGs (\S~4.4), the difference cannot be
explained by the luminosity dependence of the absorbed AGN fraction
\citep{ued03}. 
In local ULIRGs, star formation activity is concentrated at the nucleus
within $<0.5$ kpc \citep{soi00}. By contrast,
on the basis of JVLA and ALMA imaging, \citet{ruj16} reported that main
sequence SFGs at $z\sim2$ with SFR $\sim$100 \solarmass\ yr$^{-1}$ have
extended ($\sim$4 kpc diameter) star-forming regions, whereas the size
becomes more compact in more luminous SMGs (see also
\citealt{cha04,big08,ivi11} for results of radio observations). 
The
difference in the host star-forming properties between our ALMA-Chandra
sample and local ULIRGs would be related to the amount of obscuring
gas/dust around the nucleus.

We have to bear in mind, however, that the uncertainty in the column
density is often quite large due to the limited photon statistics; a few
objects (e.g., ID 1 and UDF7) tentatively classified as unobscured AGN
could be even Compton-thick AGN ($\log N_{\rm H}/ {\rm cm}^{-2} \sim
24$) within the error.
In an extreme case, only an unabsorbed scattered component
coming from outside the torus 
may be detected with Chandra in heavily Compton-thick AGN; 
subh objects would be misidentifed as (low luminosity) X-ray type-1 AGN.
In fact, according to a standard population synthesis
model of the X-ray background \citep{ued14}, the fraction of
Compton-thick AGN at $f_{\rm X} \sim 5\times10^{-17}$ \ergs\ (0.5--7
keV) is predicted to be $\sim$20\%. This corresponds to $\sim$3 out of
the 14 Chandra objects, whereas only one object is identified as Compton
thick based on the best-fit hardness ratio. Nevertheless, considering
the small number of possible additional Compton-thick AGN ($\sim$2),
main conclusions, discussed below, are not largely affected by this
uncertainty, as long as the standard X-ray background model is correct.

\subsection{Correlation between Infrared and X-ray Luminosities}

In SFGs containing AGN, the X-ray and infrared luminosities are good
indicators of the mass accretion rate onto the SMBH ($\dot{M}_{\rm BH}$)
and (dust-embedded) SFR, respectively.  Figure~3 plots the relation
between the infrared luminosity in the rest frame 8--1000 $\mu$m band
(\lir) and the intrinsic X-ray luminosity (\lx) for our ALMA-Chandra
(ASAGAO+UDF) objects. We also plot the 
Herschel-AGN sample, 
for which 
$L_{\rm IR}$ is taken from \citet{str16} and $L_{\rm X}$ is determined
from the Chandra count rates in the same way as for the ALMA-Chandra
objects. Furthermore, we plot the ALESS-AGN sample from \citet{wan13},
the NuSTAR detected U/LIRGs in the COSMOS field from \citet{mat17},
and local ULIRGs for which results from hard X-ray ($>$10 keV)
observations with NuSTAR are published. All the \lir\ values quoted in
Figure~3 are total IR luminosities including both star-forming and AGN
contributions.  In the figure, we mark the \lx -\lir\ relation expected
from pure star formation activity, based on the formula by \citet{leh10}
at SFR $> 0.4$ \solarmass\ yr$^{-1}$.  We also mark the relation
observed in Palomer-Green (PG) QSOs (i.e., AGN-dominant population,
\citealt{ten10}).

\begin{figure}
\epsscale{0.45}
\includegraphics[angle=0,scale=0.8]{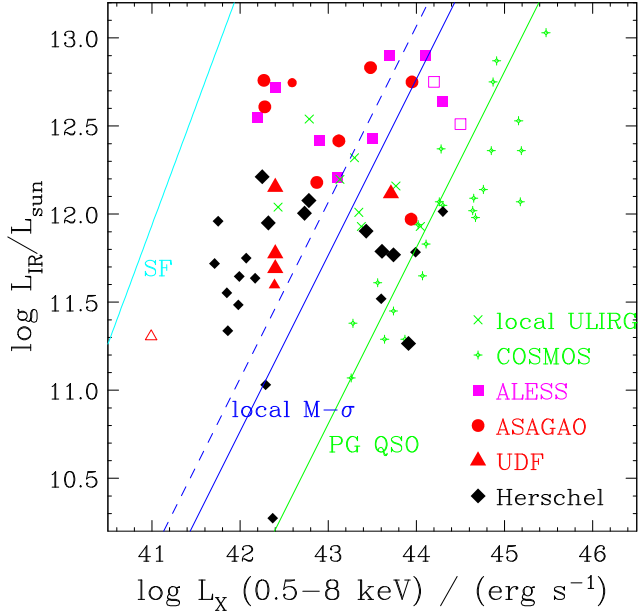}
\caption{
Relation between de-absorbed X-ray luminosity in the 
rest-frame 0.5--8 keV band (\lx ) and infrared luminosity in the 
rest-frame 8--1000 $\mu$m band (\lir) for various samples. 
Red filled circles: 
Chandra detected sources in the \ASAGAO\ sample
that are classified as AGN at $z=1.5-3$.
Red filled triangles:
Chandra detected sources in the UDF sample (D17)
that are classified as AGN at $z=1.5-3$.
Red open triangle: UDF9 (not AGN).
Magenta filled squares:
the ALESS-AGN sample \citep{wan13} at $z=1.5-3$.
Magenta open squares: those not at $z=1.5-3$.
Black diamonds: 
the Herschel-AGN sample ($z=1.5-3$).
Green diagonal crosses: 
the NuSTAR detected U/LIRGs in the COSMOS field \citep{mat17}.
Green crosses: 
local ULIRGs with available NuSTAR results \citep{ten15,oda17}.
Smaller symbols correspond to those with stellar masses of
$\log M_* / \solarmass < 10.5$.
The green solid line represent the mean value for PG QSOs \citep{ten10}.
The cyan solid line corresponds to a relation for SFGs by \citet{leh10}.
The blue solid and dashed lines are the galaxy-SMBH ``simultaneous
 evolution'' lines 
for $A=A_{\rm bul}(=200)$ (bulge only) and $A=A_{\rm tot}(=400)$
 (bulge+disk), respectively.
\label{fig3}}
\end{figure}

We find that our ALMA-Chandra AGN occupy a similar region to local
ULIRGs, having a large scatter ($>$2 dex) in the \lx /\lir\
ratio. Typical statistical errors in \lx\ and \lir\, which are
$\approx$0.2 dex and $\approx$0.1 dex, respectively (Table~1), are much
smaller than the scatter.
The possible systematic uncertainties in \lir\
(0.1--0.2 dex, Section~2) also do not affect our conclusions.
Figure~4 displays the histogram of $\log$ \lx / \lir\ for 
our ALMA-Chandra AGN 
and that including the Herschel-AGN sample.
Bimodal distribution is strongly suggested in the latter histogram.
More than half of the sources 
are distributed around a peak
centered at $\log$ \lx /\lir\ $= -3$, whereas a non-negligible
fraction of them show higher \lx\ /\lir\ ratios
consistent with AGN-dominant populations. 
While such a large variation
in the \lx -\lir\ relation is known in the local universe by combining
various samples with different selections (see e.g., Figure~8 of
\citealt{ale05}), here we find a similar \lx -\lir\ variation at
$z=1.5-3$ by directly detecting individual objects on the basis of
very deep X-ray and millimeter observations of the common survey field.
The mean IR luminosities for subsamples with $\log$ \lx\ = 42--43 and
43--44 are $\log$ \lir / \solarmass = $12.15\pm0.08$ and $12.27\pm0.10$,
respectively, which are similar to each other. This is consistent with
the result by \citet{sta15} utilizing stacking analysis that the mean IR
luminosities are fairly constant against X-ray luminosity at each
redshift. Our result is quite different from previous studies inevitably
biased for luminous AGN owning to their bright flux limits (e.g., see
Figure~2 of \citealt{wil13} for optical wide-area surveys, and
\citealt{mat17} for a hard X-ray survey with NuSTAR in the COSMOS
field), whose samples are predominantly located around the PG QSO line
in Figure~3.

\begin{figure}
\epsscale{0.45}
\includegraphics[angle=0,scale=0.8]{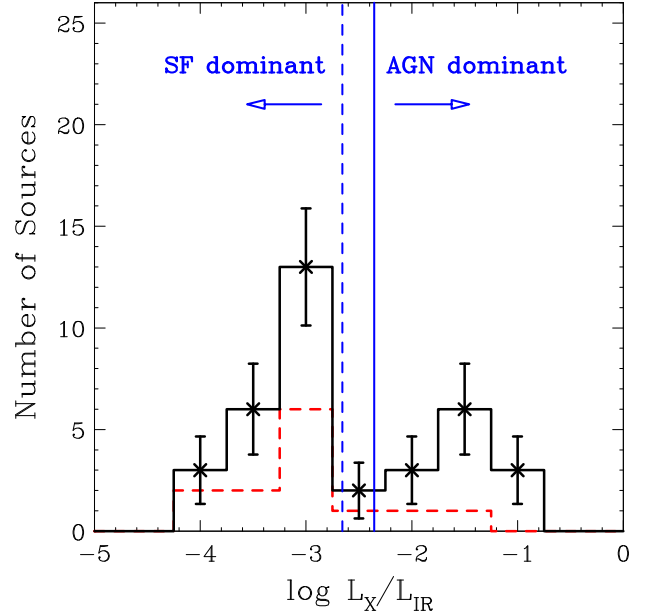}
\caption{
Histograms of $\log$ \lx / \lir\ 
for AGN in the ASAGAO and UDF samples (red)
and for AGN in the ASAGAO, UDF, and Herschel-AGN samples (black).
The error bars correspond to the 1$\sigma$ binomial uncertainties.
The blue solid and dashed lines correspond to the galaxy-SMBH 
``simultaneous evolution'' ratios 
for $A=A_{\rm bul}(=200)$ (bulge only) and $A=A_{\rm tot}(=400)$ (bulge+disk), 
respectively.
\label{fig4}}
\end{figure}

It is interesting to compare our result with the prediction of the
co-evolution scenario.
We assume the relation (for a Chabrier IMF)
\begin{equation}
{\rm SFR} = 1.09\times10^{-10} \times L_{\rm IR}/\solarlum \;\;\; (\solarmass\ {\rm yr}^{-1}).
\end{equation} 
The X-ray luminosity \lx\ can be converted
to a mass accretion rate onto the SMBH $\dot{M}_{\rm BH}$
by assuming a radiative efficiency $\eta$ and a bolometric correction 
factor $\kappa_{0.5-8}$ as 
\begin{equation}
\dot{M}_{\rm BH} = \kappa_{0.5-8} L_{\rm X} (1-\eta)/(\eta c^2),
\end{equation}
where $c$ is the light speed. 
Here we adopt $\eta=0.05$, as estimated by comparison of the local SMBH
mass density and the most updated AGN luminosity function 
\citep{ued14}\footnote{
This estimate of $\eta$ depends on $\kappa_{0.5-8}$, for which 
\citet{ued14} assumed luminosity-dependent values by \citet{hop07}. If
a constant value of $\kappa_{0.5-8}$ = 45 were adopted, we would obtain
$\eta \simeq 0.1$. This moves the blue lines in Figures~3 and 4
rightward by 0.3 dex but does not change our conclusions.}.
By analyzing the combined IR-to-mm SED of their sample, D17 estimated
that the averaged fractional AGN contribution to the IR luminosity is
$\sim$20\%. Thus, assuming that 20\% of \lir\ is the intrinsic AGN
bolometric luminosity, we can estimate an averaged bolometric correction
factor by comparing with observed X-ray luminosities. Here we ignore
X-ray undetected objects because their intrinsic X-ray luminosities
would be highly uncertain if we take into account a possibility of
heavily Compton-thick obscuration. 
Taking a luminosity-weighted average of AGN in the 
ASAGAO, UDF, and Herschel-AGN samples, we find
$\kappa_{0.5-8} \approx 45$ (or $\kappa_{2-10} \approx 70$, a
bolometric correction factor from the 2--10 keV luminosity, converted
by assuming a photon index of 1.9).

If the galaxy-SMBH
evolution is exactly simultaneous over cosmic time, we expect the relation
\begin{equation}
{\rm SFR} \times (1-R) = A \times \dot{M}_{\rm BH},
\end{equation}
where $R$ 
is the return fraction (the fraction of stellar masses that are ejected
back to interstellar medium; $R=0.41$ for a Chabrier IMF), and $A$ is the
mass ratio of stars to SMBHs in the local universe. Here we consider two
cases: $A=A_{\rm bul}$ for the stellar masses only in bulge components
and $A=A_{\rm tot}$ for the total stellar masses in both bulges and disks.
As a representative value, we use $A_{\rm bul}\sim 200$, based on the
latest calibration by \citet{kor13} at a bulge mass of $10^{11}
\solarmass$.
Adopting a total (bulge+disk) stellar mass density of $\log \rho_* /
(\solarmass {\rm Mpc}^{-3}) \simeq 8.6$ (see \citealt{mad14} and
references therein) and a SMBH mass density of $\log \rho_{\rm BH} /
(\solarmass {\rm Mpc}^{-3}) \simeq 6.0$ \citep{ued14} in the local
universe, we estimate $A_{\rm tot}\sim$400.
By the clustering analysis of $z\sim 2$ galaxies in the COSMOS field,
\citet{bet14} suggested that most of galaxies with $\log M_*/\solarmass
\sim 11$ (for a Salpeter IMF) at $z\sim2$ evolve into bulge-dominated
galaxies at $z=0$, whereas a part of SFGs with $\log M_*/\solarmass \sim
10$ at $z\sim2$ become SFGs or passive galaxies with $\log
M_*/\solarmass \sim 11$ at $z=0$.  Thus, we expect that the majority of
our objects are likely the progenitors of local bulge-dominated
galaxies, although a small fraction of them, with $\log M_*/\solarmass
\sim 10$, may end up in local disk-galaxies.  We therefore adopt
$A=A_{\rm bul}$ as the main assumption in the following discussions.

Combining equations (1), (2) and (3), we show the ``simultaneous
evolution'' relations in the \lx\ vs \lir\ plane by the blue solid
(dashed) line in Figures~3 and 4 for the bulge (total) stellar masses.
Many objects are not tightly distributed along either of these lines,
suggesting that the evolution is not simultaneous in individual
galaxies. 
The ``non-coeval'' nature is consistent with the finding by
\citet{yam09} based on K-band selected galaxies at $z\sim2$, although
they did not use far-IR/mm data to estimate the SFR. It is seen that the
majority of the $z=1.5-3$ sources are located above the lines; 
if our assumptions are correct, stars would be forming more rapidly than SMBH 
in these galaxies.
There is also a
significant fraction of objects located below this line, in which SMBHs
are growing more rapidly than the stars.
 
We note that the bolometric correction factor estimated above,
$\kappa_{0.5-8} = 45$, is larger than a nominal value for AGN with
similar X-ray luminosities ($\log L_{\rm X} \approx 42-44$),
$\kappa_{0.5-8} \approx 5-20$ \citep{rig09}. \citet{vas07} showed, however,
that the bolometric correction factor sharply correlates with Eddington
ratio rather than luminosity: the mean value of $\kappa_{2-10}$ 
rapidly increases to
$\sim$70 (or $\kappa_{0.5-8} \sim 45$) at $\lambda_{\rm
Edd} (\equiv L_{\rm bol}/L_{\rm Edd}) \simgt 0.1$. This would imply that
the Eddington ratio of our AGN is high, that is, they
contain a rapidly growing SMBH with a relatively small black hole mass.

\subsection{Correlation between X-ray Luminosity and Stellar Mass}

Figure~5 plots the correlation between X-ray luminosity and
total stellar mass for the AGN in the ASAGAO, UDF, Herschel-AGN samples.
Adopting a bolometric correction factor
$\kappa_{0.5-8} = 45$, we plot the relations that would be expected if
the ratio of the stellar mass to the black hole mass were 
$A=200$
for 3 assumed values of the Eddington ratio ($\lambda_{\rm Edd}$ = 0.01,
0.1, and 1). Apparently, the majority of the AGN are located
around the $\lambda_{\rm Edd} = 0.01$ line indicative of inefficient
SMBH accretion. This directly contradicts the previous argument that
the SMBHs in our AGN sample
have high Eddington ratios ($\lambda_{\rm Edd} > 0.1$) in
average. Time variability in the instantaneous mass accretion rate
(hence in \lx), which could produce a large \lx -to-\lir\ variation
\citep{hic14}, cannot explain this contradiction, as long as we assume
the local $M_*$-$M_{\rm BH}$ relation. This discrepancy can be solved,
however, if the black hole mass is $\sim 10$ times smaller than that
expected from the stellar mass with the local $M_*$-$M_{\rm BH}$
relation, that is, $A\sim2000$. 
In Figure~5, we also plot the ALESS-AGN sample at $z=1.5-3$; 
assuming that their Eddington ratios are also high,
they would have similarly large $A$ values on average. This
implies that these small black-hole mass systems may be young
galaxies, although we cannot find clear correlation between the
$M_*$/\lx\ ratio and the galaxy age derived from the SED fit
 (available in the ZFOURGE catalog). We leave it a future task to
pursue this issue using a larger sample.

\begin{figure}
\epsscale{0.45}
\includegraphics[angle=0,scale=0.8]{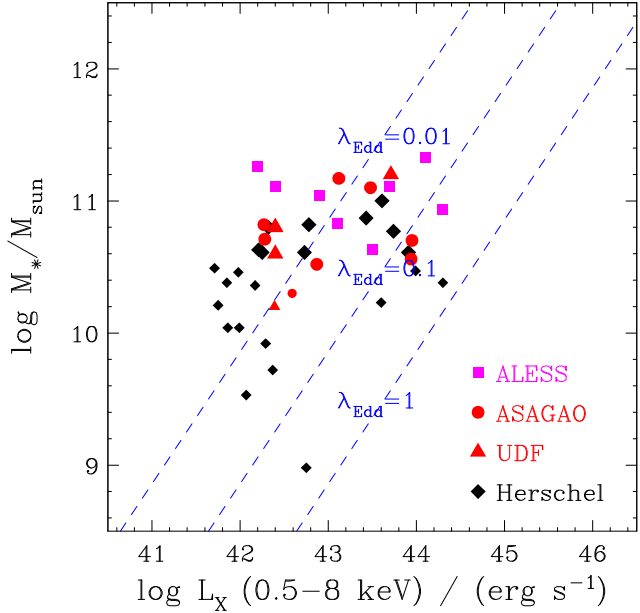}
\caption{
Relation between
de-absorbed X-ray luminosity in the rest-frame 0.5--8 keV band (\lx )
and stellar mass ($M_*$).
Red filled circles: 
Chandra detected sources in the \ASAGAO\ sample
that are classified as AGN at $z=1.5-3$.
Red filled triangles:
Chandra detected sources in the UDF sample (D17)
that are classified as AGN at $z=1.5-3$.
Magenta filled squares:
the ALESS-AGN sample \citep{wan13} at $z=1.5-3$.
Black diamonds: 
the Herschel-AGN sample ($z=1.5-3$).
Smaller symbols correspond to those with stellar masses of $\log M_* /
 \solarmass < 10.5$.
The dashed lines represent the relations expected for 3 different
Eddington ratios by assuming the local $M_*$-$M_{\rm BH}$ relation ($A=200$).
\label{fig5}}
\end{figure}

We thus infer that the majority of the ALMA/Herschel and Chandra
detected population, a representative sample of SFGs with $\log$
\lir/\solarlum $> 11$ hosting AGN at $z=1.5-3$, are in a
star-formation dominant phase and contain small SMBHs compared to their
stellar mass. This picture is in line with earlier suggestions for more
luminous (hence rare) SMGs, which are suggested to
contain small SMBHs \citep[e.g.,][]{ale05,ale08,tam10}. This may appear
to be in contrast to previous reports that luminous AGN
are in an AGN-dominant phase \citep[e.g.,][]{mat17} and have larger SMBHs
than those expected from the local $M_*$-$M_{\rm BH}$ relation (see
Figure~38 of \citealt{kor13} for a summary). We argue that the apparent
contradiction is due to selection bias for luminous AGN in such
studies. In fact, a non-negligible fraction ($\sim$20\%) 
of the whole sample, mostly X-ray luminous objects,  
are located at $\lambda_{\rm Edd} >0.1$ lines in Figure~5.
They may have SMBHs larger than those expected from
the local $M_*$-$M_{\rm BH}$ relation and truly 
be accreting with $\lambda_{\rm Edd} \sim
0.1$. 

Recalling that the majority of our sample are likely the progenitors of
local bulge-dominated galaxies, the SF dominant galaxies with small
SMBHs must experience an AGN-dominant phase later, to make the tight
$M_*$-$M_{\rm BH}$ relation at $z=0$. Indeed, such
AGN-dominant, X-ray luminous populations are present in our sample, and
even more luminous AGN were detected in wider and shallower surveys.
Our results are consistent with an evolutionary scenario that star
formation occurs first, and an AGN-dominant phase follows later
\citep[e.g.,][]{hop08,net09}, at least in objects finally evolving into
galaxies with classical bulges. 
If this is the case, the dichotomy in the \lx /\lir\ distribution
(Figure~4) would mean that the transition time from the star-formation
dominant phase to the AGN dominant one is short. 
It is interesting that, despite of
the wide diversity of populations at $z=1.5-3$, the mass-accretion rate
density and the SFR density appear to ``co-evolve'' by roughly keeping
the local $M_*$-$M_{\rm BH}$ relation \citep[e.g.,][]{mul12}. 
We suggest that
the ``co-evolution'' view is only valid when the mass accretion rate and
SFR are averaged over a cosmological timescale for an individual galaxy,
or when they are averaged for a large number of galaxies in different
evolutionary stages at a given epoch.

\section{Conclusions}

We have reported the first results from our 26 arcmin$^2$ ALMA deep
survey at the 1.2 mm wavelength on the GOODS-S field (\ASAGAO ) project,
supplemented by the deeper and narrower 1.3 mm survey by D17. This paper
focuses on the X-ray AGN properties utilizing the Chandra 7 Ms survey
data. The main conclusions are summarized as follows.

\begin{enumerate}

\item From the \ASAGAO\ survey, we detected 12 millimeter galaxies with
signal-to-noise ratio $> 5$ at a flux limit of $\approx$0.6 mJy, among
which 10 are identified by the ZFOURGE catalog. Most of them are
luminous ($\log$ \lir/\solarlum $> 11.5$ ) main-sequence SFGs at
$z=1.5-3$.

\item The AGN fraction in the ALMA detected galaxies at $z=1.5-3$ is
found to be 90$^{+8}_{-19}$\% for the ULIRGs with $\log$ \lir/\solarlum
= 12--12.8 and $57^{+23}_{-25}$\% for the LIRGs with $\log$
\lir/\solarlum = 11.5---12. The high AGN fractions among $z=1.5-3$
U/LIRGs have been revealed thanks to the much deeper X-ray flux limit
($\approx 5\times10^{-17}$ \ergs\ in the 0.5--7 keV band) than those in
previous studies.

\item There is a large variation in the \lx / \lir\ relation in the
ALMA and/or Herschel detected X-ray AGN at $z=1.5-3$. About
two thirds of them have \lx / \lir\ ratios smaller than the value
expected from the local black-hole mass to stellar mass ($M_{\rm
BH}$-$M_*$) relation. This suggests that the exactly simultaneous
co-evolution does not take place in individual galaxies.

\item If the local $M_{\rm BH}$-$M_*$ relation is assumed, the majority of
these AGN apparently show $\lambda_{\rm Edd}<0.1$. This contradicts a
large bolometric correction factor ($\kappa_{0.5-8} = 45$) estimated
from the IR SED analysis by D17 and the X-ray luminosities. We infer that a
large fraction of star-forming galaxies hosting AGN at $z=1.5-3$ have
black hole masses smaller than those expected from the local $M_{\rm
BH}$-$M_*$ relation. These results are consistent with an
evolutionary scenario that star formation occurs first, and an
AGN-dominant phase follows later, at least in objects finally evolving
into galaxies with classical bulges.

\end{enumerate}

\acknowledgments

This paper makes use of the following ALMA data:
ADS/JAO.ALMA\#2015.1.00098.S. ALMA is a partnership of ESO (representing
its member states), NSF (USA) and NINS (Japan), together with NRC
(Canada), MOST and ASIAA (Taiwan), and KASI (Republic of Korea), in
cooperation with the Republic of Chile. The Joint ALMA Observatory is
operated by ESO, AUI/NRAO and NAOJ.  The National Radio Astronomy
Observatory is a facility of the National Science Foundation operated
under cooperative agreement by Associated Universities, Inc.  Part of
this work was financially supported by Grants-in-Aid for Scientific
Research 17H06130 (YU, KK, and YT) and JP15K17604 (WR)
from the Ministry of Education,
Culture, Sports, Science and Technology (MEXT) of Japan, and by NAOJ
ALMA Scientific Research Grant Number 2017-06B. 
RJI acknowledges support from the European Research Council in the form
of the Advanced Investigator Programme, 321302, COSMICISM. 
TM and the development of CSTACK are supported by UNAM-DGAPA PAPIIT
IN104216 and CONACyT 252531.
WR is supported by the Thailand Research Fund/Office of the Higher
Education Commission Grant Number MRG6080294.

\begin{rotatetable*}
\begin{deluxetable*}{ccccccccccccccc}
\tablenum{1}
\tabletypesize{\scriptsize}
\tablecaption{Millimeter and X-ray Properties of ALMA Sources in GOODS-S
\label{tbl-1}}
\tablewidth{0pt}
\setlength{\tabcolsep}{0.03in}
\tablehead{
\colhead{ID} & \colhead{RA} & \colhead{Dec}
& \colhead{$S_{\rm 1.2mm}$} & \colhead{$S_{\rm 1.3mm}$} & \colhead{$S_{\rm 1.2mm}$}
&\colhead{$\log L_{\rm IR}$} &\colhead{$\log M_*$}
&\colhead{CID} &\colhead{$S_{\rm X}/10^{-17}$} &\colhead{$\log N_{\rm H}$} &\colhead{$\Gamma$} &\colhead{$\log L_{\rm X}$} 
&\colhead{$z$} &\colhead{AGN flag} \\
\colhead{} &\colhead{(h m s)} &\colhead{(d m s)} 
&\colhead{(mJy)} &\colhead{(mJy)} &\colhead{(mJy)} 
&\colhead{(\solarlum)} &\colhead{(\solarmass)} 
&\colhead{} &\colhead{(erg cm$^{-2}$ s$^{-1}$)} &\colhead{(cm$^{-2}$)} &\colhead{} &\colhead{(erg s$^{-1}$)} 
&\colhead{} &\colhead{}\\
\colhead{(1)} &\colhead{(2)} &\colhead{(3)} 
&\colhead{(4)}&\colhead{(5)} &\colhead{(6)} 
&\colhead{(7)} &\colhead{(8)} 
&\colhead{(9)} &\colhead{(10)} &\colhead{(11)}&\colhead{(12)} &\colhead{(13)} 
&\colhead{(14)} &\colhead{(15)}\\
}
\startdata
1 &03:32:28.51 &--27:46:58.37 &2.58$\pm$0.17 &\nodata &\nodata &12.76$\pm$0.01 &10.8 &522 & 4.7 &20.0$_{-0.0}^{+4.0}$ &1.92$_{-0.02}^{+0.00}$ &42.3$_{-0.2}^{+0.7}$ &2.38 &NNYN \\
2 &03:32:35.72 &--27:49:16.26 &2.16$\pm$0.15 &\nodata &\nodata &12.83$^{+0.04}_{-0.01}$ &11.1 &666 &39.0 &23.6$_{-0.2}^{+0.1}$ &1.9 &43.5$\pm$0.1 &2.582 &YYYN \\
3 (=UDF1) &03:32:44.04 &--27:46:35.90 &1.53$\pm$0.18 &0.92$\pm$0.08 &\nodata &12.75$\pm$0.01 &10.7$\pm$0.10 &805 &107.0 &20.0 &2.09$_{-0.11}^{+0.11}$ &44.0$\pm$0.1 &3.00 &NYYY \\
4 &03:32:47.59 &--27:44:52.39 &1.10$\pm$0.15 &\nodata &\nodata &12.61$\pm$0.01 &10.7 &852 & 8.1 &22 (fixed) &1.9 &42.3$\pm$0.2 &1.94 &NNYN \\
5 &03:32:32.90 &--27:45:40.95 &1.40$\pm$0.19 &\nodata &\nodata &11.00$\pm$0.01 &10.3 &\nodata &$<$6.9 &23 (fixed)&1.9 &$<$41.5 &0.523 &\nodata \\
6 (=UDF2) &03:32:43.53 &--27:46:39.27 &1.44$\pm$0.22 &1.00$\pm$0.09 &\nodata &12.50$^{+0.06}_{-0.01}$ &11.1$\pm$0.15 &\nodata &$<$7.5 &23 (fixed)&1.9 &$<$42.8 &2.92 &\nodata \\
7 &03:32:29.25 &--27:45:09.94 &0.89$\pm$0.17 &\nodata &\nodata &12.18$^{+0.06}_{-0.01}$ &10.5 &538 &10.5 &24.0$_{-0.6}^{+0.0}$ &1.9 &42.9$_{-0.4}^{+0.2}$ &2.01 &YYNN \\
8 (=UDF3) &03:32:38.55 &--27:46:34.52 &0.72$\pm$0.14 &0.86$\pm$0.08 &0.553$\pm$0.014 &12.75$\pm$0.01 &10.3$\pm$0.15 &718 & 4.5 &20.0$_{-0.0}^{+3.6}$ &2.44$_{-0.54}^{+0.00}$ &42.6$\pm$0.2 &2.62 &NYYN \\
9 &03:32:36.75 &--27:48:03.81 &1.05$\pm$0.25 &\nodata &\nodata &\nodata &\nodata &\nodata &$<$2.3 &\nodata &\nodata &\nodata &\nodata &\nodata \\
10 &03:32:44.60 &--27:48:36.18 &0.62$\pm$0.14 &\nodata &\nodata &11.97$\pm$0.01 &10.6 &818 &145.0 &23.1$_{-0.1}^{+0.1}$ &1.9 &43.9$\pm$0.1 &2.593 &NYYY \\
11 &03:32:49.45 &--27:49:09.21 &1.34$\pm$0.31 &\nodata &\nodata &\nodata &\nodata &\nodata &$<$5.7 &\nodata &\nodata &\nodata &\nodata &\nodata \\
12 &03:32:31.47 &--27:46:23.38 &1.11$\pm$0.28 &\nodata &\nodata &12.42$\pm$0.01 &11.2 &587 &26.3 &23.3$_{-0.2}^{+0.2}$ &1.9 &43.1$\pm$0.1 &2.225 &YYYN \\
UDF4 &03:32:41.01 &--27:46:31.58 &\nodata &0.30$\pm$0.05 &\nodata &11.92$\pm$0.02 &10.5$\pm$0.15 &\nodata &$<$4.6 &23 (fixed)&1.9 &$<$42.5 &2.43 &\nodata \\
UDF5 &03:32:36.95 &--27:47:27.13 &\nodata &0.31$\pm$0.05 &\nodata &11.95$\pm$0.03 &10.4$\pm$0.15 &\nodata &$<$3.9 &23 (fixed)&1.9 &$<$42.2 &1.759 &\nodata \\
UDF6 &03:32:34.43 &--27:46:59.77 &\nodata &0.24$\pm$0.05 &\nodata &11.88$\pm$0.06 &10.5$\pm$0.10 &\nodata &$<$5.1 &23 (fixed)&1.9 &$<$42.1 &1.411 &\nodata \\
UDF7 &03:32:43.32 &--27:46:46.91 &\nodata &0.23$\pm$0.05 &\nodata &11.69$\pm$0.18 &10.6$\pm$0.10 &797 & 6.6 &20.0$_{-0.0}^{+4.0}$ &1.9 &42.4$_{-0.2}^{+0.7}$ &2.59 &NNYN \\
UDF8 &03:32:39.74 &--27:46:11.63 &\nodata &0.21$\pm$0.05 &0.223$\pm$0.022 &12.12$\pm$0.27 &11.2$\pm$0.15 &748 &284.0 &22.6$_{-0.1}^{+0.1}$ &1.9 &43.7$\pm$0.1 &1.552 &NYYY \\
UDF9 &03:32:43.42 &--27:46:34.46 &\nodata &0.20$\pm$0.04 &\nodata &11.31$\pm$0.48 &10.0$\pm$0.10 &799 & 4.2 &22 (fixed) &1.9 &41.0$\pm$0.2 &0.667 &NNNN \\
UDF10 &03:32:40.75 &--27:47:49.09 &\nodata &0.18$\pm$0.05 &\nodata &11.60$\pm$0.22 &10.2$\pm$0.15 &756 & 2.5 &20.0$_{-0.0}^{+3.0}$ &3.00$_{-1.10}^{+0.00}$ &42.4$\pm$0.2 &2.086 &NNYN \\
UDF11 &03:32:40.06 &--27:47:55.82 &\nodata &0.19$\pm$0.05 &\nodata &12.15$\pm$0.26 &10.8$\pm$0.10 &751 & 9.3 &21.8$_{-1.9}^{+1.0}$ &1.90$_{-0.00}^{+0.42}$ &42.4$_{-0.1}^{+0.2}$ &1.996 &NNYN \\
UDF12 &03:32:41.28 &--27:47:42.61 &\nodata &0.15$\pm$0.04 &\nodata &11.51$\pm$0.17 & 9.6$\pm$0.15 &\nodata &$<$2.0 &23 (fixed)&1.9 &$<$42.7 &5.000 &\nodata \\
UDF13 &03:32:35.09 &--27:46:47.78 &\nodata &0.17$\pm$0.04 &\nodata &11.78$\pm$0.12 &10.8$\pm$0.10 &655 & 4.7 &20.0$_{-0.0}^{+3.8}$ &2.07$_{-0.17}^{+0.00}$ &42.4$_{-0.2}^{+0.5}$ &2.497 &NNYN \\
UDF14 &03:32:40.96 &--27:46:55.34 &\nodata &0.16$\pm$0.04 &\nodata &11.59$\pm$0.17 & 9.7$\pm$0.10 &\nodata &$<$3.3 &23 (fixed)&1.9 &$<$41.5 &0.769 &\nodata \\
UDF15 &03:32:35.75 &--27:46:54.98 &\nodata &0.17$\pm$0.05 &\nodata &11.52$\pm$0.31 & 9.9$\pm$0.15 &\nodata &$<$1.8 &23 (fixed)&1.9 &$<$41.8 &1.721 &\nodata \\
UDF16 &03:32:42.37 &--27:47:07.79 &\nodata &0.15$\pm$0.04 &\nodata &11.55$\pm$0.20 &10.9$\pm$0.10 &\nodata &$<$4.0 &23 (fixed)&1.9 &$<$42.0 &1.314 &\nodata
\enddata
\tablecomments{
(1) ALMA source ID (those with ``UDF'' correspond to the D17 sources), 
(2)(3) ALMA source position (J2000), 
(4) ALMA integrated flux-density and 1$\sigma$ error at 1.2 mm derived
 from the ASAGAO survey,
(5) those at 1.3 mm derived from the UDF survey (D17),
(6) those at 1.2 mm derived from the ASPECS \citep{ara16},
(7) infrared luminosity in the rest 8--1000 $\mu$m band 
in units of solar luminosity (taken from D17 for the UDF sample
and based on our SED fit for the ASAGAO sample) and its 1$\sigma$ error,
(8) stellar mass in units of solar mass (taken from D17 for the UDF
sample and IDs. 3, 6, and 8, and from \citet{str16} for the rest),
(9) Chandra source ID in \citet{luo17},
(10) observed X-ray flux (or 90\% confidence upper limit) in the 0.5--7 keV band converted from a count
rate in the 0.5--7, 0.5--2, or 2--7 keV band
with the apparent photon index $\Gamma_{\rm eff}$, 
(11) X-ray absorption hydrogen column density,
(12) intrinsic photon index, 
(13) absorption-corrected X-ray luminosity (or 90\% confidence upper limit) in the rest-frame 0.5--8 keV band
(14) adopted redshift (after \citealt{str16} or D17; 
3 decimal digits for spectroscopic redshifts and 2 for photometric redshifts),
(15) AGN flags for the criteria I through IV. 
}
\end{deluxetable*}
\end{rotatetable*}

\end{document}